\documentclass[sigconf]{acmart}
\AtBeginDocument{%
  \providecommand\BibTeX{{%
    \normalfont B\kern-0.5em{\scshape i\kern-0.25em b}\kern-0.8em\TeX}}}

\copyrightyear{2023}
\acmYear{2023}
\setcopyright{acmlicensed}\acmConference[UIST '23]{The 36th Annual ACM Symposium on User Interface Software and Technology}{October 29-November 1, 2023}{San Francisco, CA, USA}
\acmBooktitle{The 36th Annual ACM Symposium on User Interface Software and Technology (UIST '23), October 29-November 1, 2023, San Francisco, CA, USA}
\acmPrice{15.00}
\acmDOI{10.1145/3586183.3606786}
\acmISBN{979-8-4007-0132-0/23/10}

\usepackage{xcolor}
\definecolor{ent}{HTML}{606266}
\definecolor{col}{HTML}{DAA520}

\usepackage{xspace}
\usepackage{lipsum}
\usepackage{comment}
\usepackage{hyperref}
\usepackage{enumerate}
\usepackage[inline]{enumitem}
\usepackage[noend]{algorithmic}
\usepackage{fontawesome}
\usepackage{cuted}
\usepackage{amsmath}

\newcommand{\xb}[1]{{\color{black} #1}}

\newcommand{\pzh}[1]{{\color{black} #1}}
\newcommand{\xingbo}[1]{{\color{black} #1}}

\newcommand{\ie}{\textit{i.e.,} }
\newcommand{\eg}{\textit{e.g.,} }
\newcommand{\etal}{\textit{et al.} }

\newcommand{\name}{{\textit{Storyfier}}}

\usepackage{algorithm}
\usepackage{algorithmic}

\usepackage{lineno}

\begin{document}

\title{
Storyfier: Exploring Vocabulary Learning Support with Text Generation Models 
}

\author{Zhenhui Peng}
\authornote{Both authors contributed equally to this research.}
\email{pengzhh29@mail.sysu.edu.cn}
\orcid{0000-0002-5700-3136}
\affiliation{%
  \institution{Sun Yat-sen University}
  \city{Zhuhai}
  \country{China}
}

\author{Xingbo Wang}
\email{xwangeg@cse.ust.hk}
\orcid{0000-0001-5693-1128}
\authornotemark[1]
\affiliation{%
  \institution{The Hong Kong University of Science and Technology}
  \city{Hong Kong}
  \country{China}
}

\author{Qiushi Han}
\orcid{0009-0003-5455-579X}
\affiliation{%
  \institution{Sun Yat-sen University}
  \city{Zhuhai}
  \country{China}
  }
\email{hanqsh@mail2.sysu.edu.cn}

\author{Junkai Zhu}
\orcid{0009-0001-8654-7829}
\affiliation{%
  \institution{Guangdong Polytechnic of Industry \& Commerce}
  \city{Guangzhou}
  \country{China}
}
\email{zhujunkai@hotmail.com}

\author{Xiaojuan Ma}
\orcid{0000-0002-9847-7784}
\affiliation{%
 \institution{The Hong Kong University of Science and Technology}
  \city{Hong Kong}
  \country{China}}
\email{mxj@cse.ust.hk}

\author{Huamin Qu}
\orcid{0000-0002-3344-9694}
\affiliation{%
 \institution{The Hong Kong University of Science and Technology}
  \city{Hong Kong}
  \country{China}}
\email{huamin@cse.ust.hk}

\renewcommand{\shortauthors}{Zhenhui Peng and Xingbo Wang, et al.}

\begin{abstract} 
Vocabulary learning support tools have widely exploited existing materials, e.g., stories or video clips, as contexts to help users memorize each target word. However, these tools could not provide a coherent context for any target words of learners' interests, and they seldom help practice word usage. In this paper, we work with teachers and students to iteratively develop Storyfier, which leverages text generation models to enable learners to read a generated story that covers any target words, conduct a story cloze test, and use these words to write a new story with adaptive AI assistance. Our within-subjects study (N=28) shows that learners generally favor the generated stories for connecting target words and writing assistance for easing their learning workload. However, in the read-cloze-write learning sessions, participants using Storyfier perform worse in recalling and using target words than learning with a baseline tool without our AI features. We discuss insights into supporting learning tasks with generative models.
\end{abstract}





\maketitle
\section{Introduction}
\pzh{
Learning vocabulary in meaningful contexts, such as stories and images in language learning textbooks, and video clips from movies, is a common and effective practice as it enables deep and active processing of vocabulary (\eg word associations, logic) \cite{oxford1994second}.
\xingbo{Many existing vocabulary learning systems like VocabEncounter \cite{VocabEncounter} and Smart Subtitles \cite{kovacs2014smart} have exploited a variety of materials to establish the contexts for words. 
These systems have demonstrated that the provided contexts can enhance vocabulary memorization \cite{VocabEncounter,kovacs2014smart}.} 
}

However, these systems may fall short in two aspects.
First, they largely leverage existing materials and could not provide a meaningful context for any set of target words that users wish to learn. 
In other words, previous systems lack the flexibility to offer a story, an article, or a video clip that covers the target words that teachers or learners specify. 
These coherent contexts that connect target words may make a difference to vocabulary learners.
As suggested by Gu \etal~\cite{vocabulary_learning_strategies}, learning vocabulary in batches under coherent contexts could facilitate recalls of a larger amount of words compared to learning vocabulary in isolation. 
Second, previous systems primarily focus on helping users to understand and memorize the meanings of target words via meaning-focused input learning activities, \eg reading and listening, that use language receptively \cite{nation2009teaching}. 
Few systems facilitate learners to master the usage of learned words via productive and fluency development tasks (\eg writing and speaking) -- typical activities that could help master the meanings and usage of target words in traditional courses \cite{nation2009teaching}. 
In offline courses, teachers can provide in-situ adaptive support like hints on word usage during these learning activities; however, this is often unavailable to individual learners outside classrooms.

\xingbo{
In this work, we utilize stories 
generated by large language models (LLMs) as meaningful contexts that cover any target words and provide adaptive assistance in word usage practice. 
Our focus is motivated, on one hand, by the prevalent use of stories in language learning textbooks, and the proven efficacy of story-based learning in various scenarios such as programming ~\cite{dietz2021storycoder, dietz2023visual, suh2022codetoon}, parent-child storytelling ~\cite{zhang2022storybuddy}, and children's visual storytelling~\cite{zhang2022storydrawer}. 
On the other hand, LLMs can generate fluent and relevant texts given user specifications such as keywords, which have been used to support the writings of emails \cite{goodman2022lampost}, articles or fiction \cite{dang2022beyond}, and poems \cite{shao2021sentiment, van2020automatic}. 
However, little work, if any, has explored LLMs for story-based vocabulary learning where users should spend effort in mastering target words' meanings and usage.
Questions arise such as 1) whether and how LLMs can generate the meaningful context of any target word set for vocabulary learning, 2) if so, what vocabulary learning activities can these generative models support, and 3) how would the support from generative models impact the users' vocabulary learning outcome and experience. 
}

\pzh{To this end,} we seek to provide insights into these questions by designing, developing, and evaluating an AI-generated story-based vocabulary learning system, \name{}, that can provide meaningful story contexts and adaptive assistance for learning any set of target words. 
Here, we choose English as the target language to learn and target ESL (English-as-the-Second-Language) Chinese learners, \eg high-school or university students in China. 
We take an iterative design approach with insights from educational literature and the involvement of teachers, learners, and HCI researchers in this process. 
We first fine-tune a text-generation model on a short-story corpus and validate its capability in producing meaningful story context given a set of target CET-4 English words \footnote{Short for College English Test Band 4, a mandatory test for acquiring bachelor degrees in China.}. 
We then present this model to three English teachers and five experienced ESL learners in an interview study to explore possible learning activities that \name{} can support. 
Based on the insights from the interviews and educational literature, we develop a \name{} prototype that supports three types of vocabulary learning activities: 1) \textit{reading} an AI-generated story with target words,
2) solving story \textit{cloze} tests on target words \pzh{(\ie fill blanks of the generated story by using target words)},
and 3) \textit{writing} a story using target words with the AI models by turns. 
We seek feedback on \name{}'s design and refine it via a user study with twelve ESL learners and two co-design workshops with the three English teachers mentioned above and four HCI researchers. 

We conduct a $2 \times 2$ within-subjects study with 28 university students to evaluate the impact of \name{}'s AI functions (with vs. without generative models) and learning activity (read-only vs. read-cloze-write) on the learning outcome and experience. 
The results show that in the read-only learning sessions, the generative stories do not help to improve learning gains in recalling target words' meanings and mastering their usage. 
In read-cloze-write learning sessions, participants with generated stories and AI assistance perform even worse compared to the condition without the generative models. 
However, most participants still indicate their preferences on \name{}'s generated stories for connecting target words and its writing assistance for reducing learning workload. 
Based on our findings, we highlight the value of generative models in offering meaningful materials and enjoyable experience for learning tasks. 
We also urge future AI-supported learning tools to ensure users to spend the necessary effort in their learning tasks. 

Our work makes three contributions. 
First, we present a vocabulary learning system \name{} that facilitates users to master the meanings and usage of any target English words via AI-generated stories and writing assistance.
Second, our design and evaluation of \name{} provide first-hand findings on the feasibility, effectiveness, and user experience of applying generative models to vocabulary learning. 
Third, we offer insights and design considerations of leveraging generative models to support learning tasks. 
\section{Related Work}
To situate our work, we start by reviewing the pedagogical strategies and activities for vocabulary learning. 
We then discuss previous vocabulary learning support systems.
Lastly, we introduce related textual story-generation techniques that enable us to achieve the envisioned \name{}.   

\subsection{Pedagogical Strategies and Activities For Vocabulary Learning}
\label{subsection_2.1}
According to the amount of context information used, vocabulary learning strategies can be categorized as decontextualized, partially-contextualized, and fully-contextualized~\cite{oxford1990vocabulary, gu1996vocabulary}.
Decontexutualized techniques, including using word lists in alphabetical order or by part of speech, flashcards, dictionary, focus on learning isolated words without meaningful contexts. 
For example, \textit{dictionary} provides detailed instructions on grammar, pronunciation, and brief usage examples. However, improper use of dictionary, \eg checking every word's meaning during reading and failing to associate it with the current context, would result in poor learning outcomes~\cite{swaffar1988readers}.
In other words, decontextualized techniques may not aid long-term vocabulary retention and practical word usage~\cite{vocabulary_learning_strategies, taka2008vocabulary}.  

Educators have argued that vocabulary is better learned through contextualized learning activities~\cite{gu1996vocabulary}.
Partially-contextualized techniques provide a certain amount of context information (\eg word association).
For instance, \textit{Word grouping} organizes words according to different criteria, such as (dis)similarity and topic.
\textit{Concept association} (or ``elaboration'') constructs connections between new words and some familiar contexts, such as previously learned words, personal experience, or knowledge in learners' memory~\cite{carrell1984schema}.
Besides, \textit{keyword techniques}~\cite{pressley1982mnemonic} link words with visual~\cite{bower1970analysis} or aural~\cite{dunn1972practical} objects to improve vocabulary memorization.
\textbf{Fully contextualized techniques }associate words in fully authentic communication contexts and connect them with a meaningful flow (\eg logic), which are considered the peak of L2 vocabulary learning techniques~\cite{oxford1990vocabulary}.
They use existing newspapers, articles, magazines, and novels as learning material. 
The most common activities are reading or listening to the stories in contextual inference tasks (\eg cloze test)~\cite{gu1996vocabulary}.
Speaking and writing practices are regarded as the more effective but also challenging activities, which require turning receptive vocabulary knowledge into productive use in communication contexts~\cite{oxford1989language}.
Nevertheless, contextualized methods are demanding and complex for individual learners
and are usually adopted by teachers in classroom activities~\cite{schmitt2008instructed, gu1996vocabulary}. 

Regarding the learning activities in a traditional language course, Paul Nation, 
suggested that there should be roughly equal amounts of time given to each of the following four strands \cite{nation2007four}. 
The \textbf{meaning-focused input} strand involves learning through listening and reading -- using language receptively. This strand mainly focuses on understanding what they listen to and read, e..g, stories, TVs, films, conversations, and so on.
The \textbf{meaning-focused output} strand involves learning through speaking and writing -- using language productively. Typical activities in this strand include talking in conversations, writing a letter or a note, keeping a diary, telling a story, etc. 
The \textbf{language-focused learning}  strand involves the deliberate learning of language features such as pronunciation, spelling, vocabulary, grammar, and
discourse. 
Lastly, the \textbf{fluency development strand} should involve four skills of listening, speaking, reading, and writing. 
In this strand, learners are helped to make the best use of what they have already known in typical activities like ten-minutes writing and listening to easy stories. 
These four strands can fit together in many different ways \cite{nation2007four,nation2009teaching}. 
For example, a group collaborative writing activity in the high-school can combine the meaning-focused output and language-focused learning strands, if the output written work deliberately focuses on the vocabulary and grammar \cite{nation2007four}. 

Our work is motivated by the benefits of fully contextualized strategies and gets inspired by the four strands of activities for vocabulary learning. 
We use textual short stories as contextualized vocabulary learning materials. 
We support individual vocabulary learners with a proper integration of the four strands of learning activities 
based on the story contexts.

\subsection{Vocabulary Learning Support Systems}
Researchers have proposed various approaches and systems to support vocabulary learning. 
For rote learning, a bunch of work manage to model users' memory cycles and plan the target words with proper difficulty level and repetition frequency~\cite{nioche2021improving, chen2008personalized, zeng2011interactive,how_people_learn_and_forget}. 
As for our focused contextual learning, previous vocabulary learning systems have exploited materials in different mediums, such as images \cite{tangworakitthaworn2019image}, physical locations \cite{yamamoto2019findo}, textual articles in webpage \cite{VocabEncounter}, subtitles of videos \cite{kovacs2014smart, huang2012ubiquitous, sakunkoo2013gliflix}, and augmented/virtual reality \cite{santos2016augmented,pu2018development,hsieh2006interaction}.
For instance, FinDo~\cite{yamamoto2019findo} is a mobile application that helps users understand the vocabulary about the surrounding objects with the contexts of users' current locations.
Tangworakitthaworn \etal~\cite{tangworakitthaworn2019image} used image processing techniques to extract visual objects in photos and matched them with the target vocabulary.
VocabEncounter~\cite{VocabEncounter} encloses target vocabulary into reading materials to facilitate micro learning in daily life.
Smart Subtitles \cite{kovacs2014smart} equips video subtitles with features like vocabulary definitions on hover and dialog-based video navigation. 


However, these systems largely make use of existing materials as contexts, which may not be able to provide a meaningful flow that covers any set of target words -- a requirement of fully contextualized learning \cite{oxford1990vocabulary}. 
We seek to mitigate this constraint by generating a short story for any target word set. 
\pzh{Our decision to use stories as the context for words is inspired by their common usage in language learning textbooks and the proven efficacy of story-based learning in other scenarios~\cite{dietz2021storycoder, dietz2023visual, suh2022codetoon,zhang2022storydrawer,zhang2022storybuddy}.}
Further, previous vocabulary learning support systems mainly focus on supporting meaning-focused input activities that aim at understanding the words' meanings. 
Our work further supports other types of learning activities that help to master the usage of target words. 


\subsection{Textual Story Generation Techniques}

Recent advances in textual story generation offer potentials to support vocabulary learners with meaningful contexts that cover any word set and offer in-situ learning support. 
The textual story generation techniques aim at generating coherent and fluent narratives or ideas based on simple user inputs, such as a title~\cite{mostafazadeh-etal-2016-corpus} and prompts~\cite{fan2018hierarchical}.
Early computational work adopts symbolic approaches~\cite{porteous2009controlling, riedl2010narrative, ware2013computational} that first select a sequence of characters and actions according to  aesthetic, narrative conflicts, and logic, and then create a story with pre-defined templates. 
Another approach is case-based reasoning~\cite{gervas2004story,turner2014creative}, which extracts the story plots of existing stories and adapts them to new contexts. 
Yet, these methods are restricted by predefined story domains and styles. 
Recent story generation methods mainly adopt sequence-to-sequence language models~\cite{fan2018hierarchical, yao2019plan, harrison2017toward, martin2018event}, which can learn complex and implicit relationships among story plots.
Particularly, transformer-based models~\cite{see2019massively, keskar2019ctrl, brown2020language} are able to produce incredibly fluent texts after training on a large language corpus.
These models can be finetuned to support downstream applications like writing assistants~\cite{calderwood2020novelists, buschek2021impact} and health consultation~\cite{wang2021evaluation}. 


To generate stories with desired properties (\eg keywords, topic, styles), researchers apply techniques like decoding strategies, prompt controls, and finetuning to build controllable language models.
Decoding strategies aim to restrict and influence the sampling process of generation to change the features of output texts. These features can describe the user preferences and are modeled by heuristics~\cite{ghazvininejad2017hafez}, supervised signals~\cite{holtzman2018learning}, and reinforcement learning~\cite{li2017learning}.
Prompt controls use natural language (\eg ``translate to English'') to elicit desired contents \cite{radford2019language, brown2020language,shin2020autoprompt, jiang2020can, li2021prefix, liu2021gpt, lester2021power}. 
Finetuning methods investigate effective conditional training 
based on key words~\cite{fan2018hierarchical}, story valence~\cite{peng2018towards}, character fortune~\cite{chung2022talebrush}, control codes~\cite{keskar2019ctrl} (\eg, topic, sentiment), and simpler attribute models~\cite{dathathri2019plug, krause2020gedi}.

\pzh{
Recent intelligent systems have explored the usage of text generation techniques in a variety of scenarios, such as creative writing \cite{chi23_journalistic_ideation,chi23_co-writing}, AI-mediated communication \cite{AI_mediated_communication}, and health intervention \cite{chi23_health_intervention}. 
In the story-based learning scenario, 
StoryBuddy~\cite{zhang2022storybuddy} assists parents-children storytelling via a question-answer generation model, which consists of a rule-based answer generation module, a BART-based question generation module, and a ranking module.
It can help parents create a storytelling bot that can tell stories, ask children questions, and provide feedback ~\cite{zhang2022storybuddy}.
}
\xingbo{
However, these studies present a different focus compared to ours. We specifically investigate the use of and interaction with text generation models for story-based vocabulary learning.}


In this paper, we first customize and evaluate a controllable language model for generating meaningful stories that cover given target word set. 
We then explore what vocabulary learning activities that this model can support with teachers and students. 

\section{Phase 1: Feasibility and Supported Activities of Story Generation Models for Vocabulary Learning}

\begin{figure}
\includegraphics[width=0.45\textwidth]{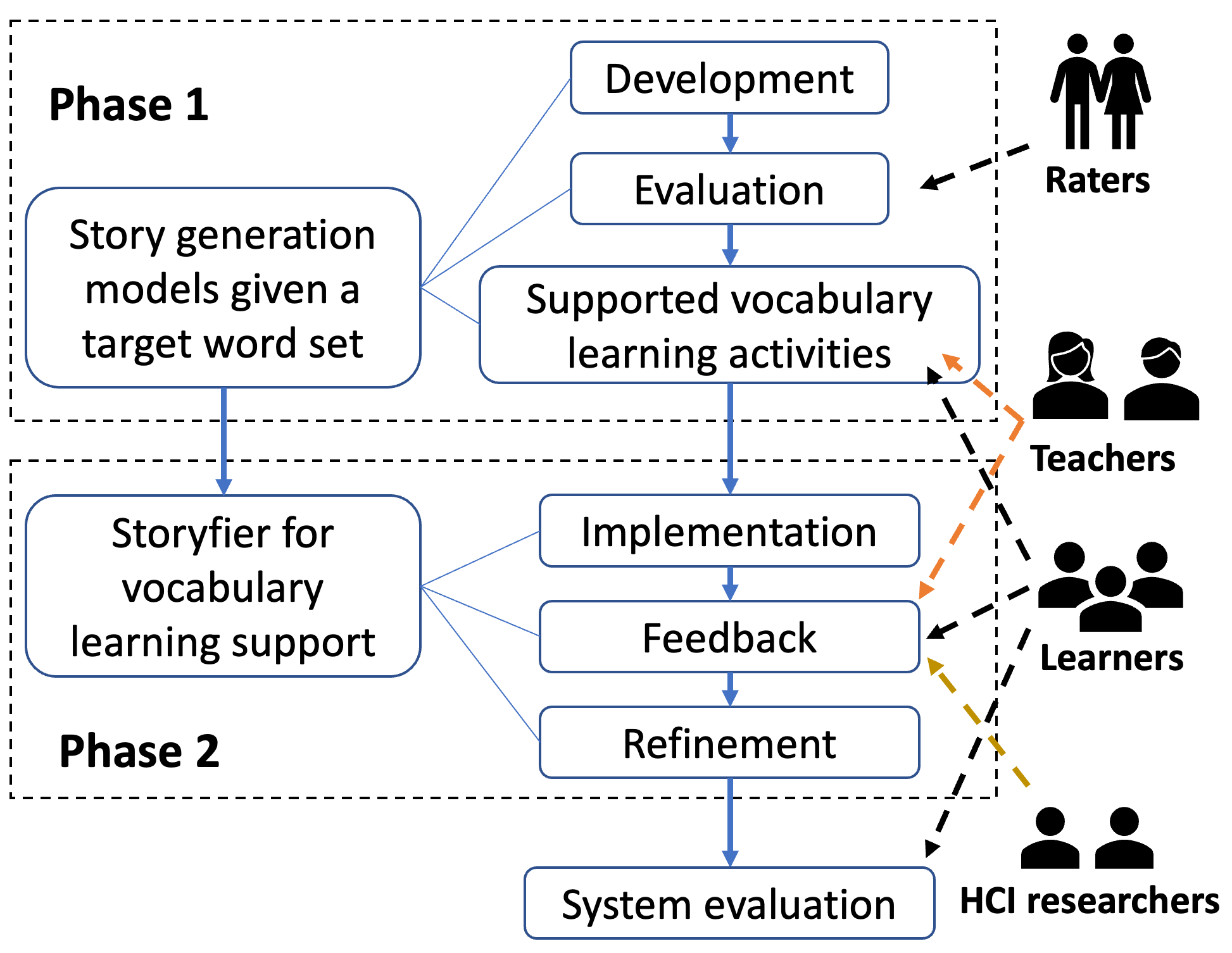}
    \caption{Our two-phases design and development process of \name{} with teachers, learners, and HCI researchers.}
    \label{fig:two_phases}
    \Description{Figure 1 shows our two-phases design and development process of Storyfier with teachers, learners, and HCI researchers. Phase 1 contains Story generation models given a target word set, which consists of Development, Evaluation (with human Raters), and Supported vocabulary learning activities (with Teachers and Learners). The models and learning activities in Phase 1 feed into Phase 2. Phase 2 contains Storyfier for vocabulary learning support, which consists of Implementation, Feedback (with Teachers, Learners, and HCI researchers), and Refinement. The Storyfier Phase 2 feeds into System evaluation (with Learners).}
\end{figure}

To help individual learners to master the meaning and usage of any target word sets, we design and develop \name{} via a two-phase process (\autoref{fig:two_phases}). 
In this section, we present the first phase in which we 1) develop a story generation model, 2) validate its feasibility for providing meaningful contexts for vocabulary learning, and 3) explore what vocabulary learning activities this model could support. 


\subsection{Developing Story Generation Models}


Given the potential benefits of a meaningful story for learning a batch of words ~\cite{vocabulary_learning_strategies}, we first seek to develop a controllable language model that can generate stories with given target word sets. 
Here, we target the vocabulary pool (4,827 in total) required by the College English Test Band 4 (CET-4), a mandatory national test for Chinese university students to obtain bachelor degrees.



\subsubsection{Dataset}
We choose ROCStory corpus~\cite{mostafazadeh-etal-2016-corpus} to contextualize CET-4 words and build story generation models. 
ROCStory collects over 100,000 five-sentence commonsense human-written stories (\autoref{table.rocstory}\footnote{Readability is measured by \href{https://en.wikipedia.org/wiki/Flesch-Kincaid_readability_tests}{Flesch Reading-Ease}, and CET-4 is 34.23 on average.}).
The simple and short story form could help learners easily understand the story flow and mitigate diversion from vocabulary learning to story comprehension.
\xingbo{The simplicity of the story structures and logic is also appreciated by English teachers who participate in the later studies (\autoref{sec:interview_teachers}).}
Though these stories are short, they are created by various human workers and have passed qualification tests to ensure story quality and creativity.
In addition, these stories have causal and temporal commonsense relationships between story sentences and cover a wide range of everyday topics, such as movie, school, birthday, and music. 
Therefore, if there is a story that covers a set of target words, learners can easily associate a group of words with a common topic following a meaningful logic flow. 
With this dataset, we aim to develop a model that can generate meaningful stories like those in ROCStory given any set of target words in the CET-4 pool. 


\begin{table}[t]
\caption{The statistics of ROCStory dataset.}
\label{table.rocstory}
\begin{tabular}{ll}
\hline
\textbf{Attributes}     & \textbf{Values} \\ \hline
\# of stories           &      101,661           \\
\# of words             &          4,640,319       \\
Average story length         &    45.65             \\
Average sentence length         &      7.80           \\
Average readability    &         57.14       \\
Coverage of CET-4 words &        89.52\%         \\ \hline
\end{tabular}
\end{table}

\begin{figure*}[t]
    \includegraphics[width=\textwidth]{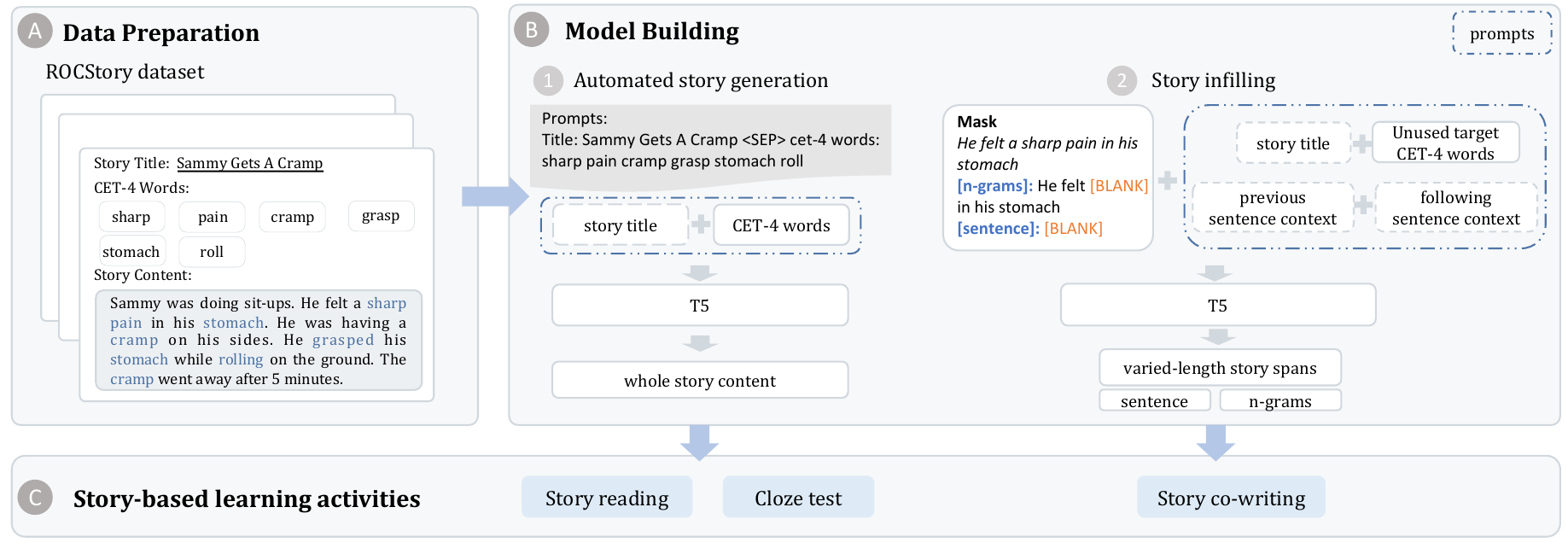}
    \caption{
    The technical framework of \name{}. We mainly adopt prompt-based fine-tuning strategies to build story generation models. (A) We derive CET-4 words from the stories in ROCStory dataset. (B) We finetune T5 language models to 1) generate a story given a CET-4 word set with or without a title (presented in section 3.2) and 2) infill a sentence or n-grams given preceding and following sentences, unused target words, and story title (if any) (\autoref{section_4.2.1}). 
    (C) We apply the models to support three kinds of story-based learning activities. 
    }
    \label{fig:algorithm}
    \Description{Figure 2 shows the technical framework of Storyfier. Three components: Data Preparation, Model Building, and Story-based learning activities. Data Preparation describes our pre-processing of ROCStory dataset. It feeds into Model Building, which consists of Automated story generation and Story infilling. These feed into Story-based learning activities, which includes Story reading, Cloze test, and Story co-writing. }
\end{figure*}

\subsubsection{Data Preprocessing and Model Building}
\label{subsec.story_generation}
\autoref{fig:algorithm} summarizes our data preprocessing and model building procedure. 
Specifically, we follow recent story generation techniques \cite{fan2018hierarchical,peng2018towards,chung2022talebrush} and formulate the problem as a sequence-to-sequence translation task. 
We first segment the stories into titles\footnote{For the stories without titles, we represent their title features as ``no title''.} and sentences. 
Then, using the CET-4 word list, we identify the occurrences of these words in each sentence of every story and sort them chronologically. This leads to the creation of a set of \{\textit{story title, target words, story sentences}\} tuples (\autoref{fig:algorithm}A).
\xingbo{For story generation, we leverage a state-of-the-art open-source language model T5~\cite{raffel2019exploring} as the base model. 
Our decision is made based on two reasons.
First, T5 exhibits impressive performance across various NLP tasks (\eg text generation and classification), which can be attributed to its unified text-to-text framework and its pretraining on a large language corpus. 
Moreover, it is freely available and adaptable to our application scenario compared to other impressive but closed-source language models (\eg GPT-3~\cite{brown2020language} and GPT-4~\cite{openai2023gpt4}).}

Then, we adopt a prompt-based approach to finetune and steer the model generation process to 
learn the mappings between target English words and a story (\autoref{fig:algorithm}B1). 
We formulate the input prompt as the concatenation of the \textit{story title} and \textit{target words}
derived from story tuples of the processed dataset. 
According to our experiments, the \texttt{title} imposes a high-level control of story relevance and leads to faster and better convergence compared to training without the \texttt{title} signal.
We finetune the pretrained T5-large model offered in the HuggingFace on our dataset using Adam optimization algorithm with a 0.0001 learning rate. 
The training process lasts for five epochs and has a 0.9857 cross-entropy loss. 

\xingbo{After training, our model can generate complete stories rather than isolated sentences, thus creating meaningful contexts for the target words across multiple sentences. For instance, when given the words ``athlete'', ``avid'', and ``frequently'' (as shown in  \autoref{fig:system}), the model begins a narrative about an avid athlete who frequently participates in marathons.}

\subsection{Evaluating the Quality of Generative Stories}
While our model can generate a story given any word set with or without a title, at this stage, we would like to compare the quality of machine-generated and human-written stories in the corpus which cover the same target word set. 
\pzh{
This evaluation aims at validating if the generated stories were competent for vocabulary learning support.
We will assess the perceived quality and helpfulness of generated stories given any target words in our later interviews with teachers (\autoref{sec:interview_teachers}) and experiments with learners (\autoref{subsection:6.3.2}).} 
Following prior work~\cite{chung2022talebrush,fan2018hierarchical,peng2018towards}, we conduct technical and human evaluations. 
We sample 20 stories from the ROCStory dataset with varied difficulty levels (\ie word frequency) of contained CET-4 words. 
For each human-written story, we use our trained language models to generate a machine version based on the story title and contained CET-4 words.
Thereafter, we create 20 human-machine story pairs (40 stories in total).

\subsubsection{Technical Evaluation}
We assess the story content from grammatical accuracy, lexical diversity (\ie number of unique words, and type-token ratio: number of unique words/total number of words), and lexical coherence (\ie trigram repetition, and sentence coherence \footnote{Cosine similarities (range 0-1) between sentence embeddings using \href{https://www.sbert.net/}{\textit{sbert}}.}: average semantic similarities between sentences) ~\cite{roemmele2017evaluating, lapata2005automatic, goldfarb2020content}.
As shown in \autoref{table.lexical_eval}, 
both the machine-generated and human-written stories have no grammar issues.
Moreover, the machine performance is commensurate with the human in terms of lexical diversity and coherence, as indicated by close scores of type-token ratio, trigram repetition, and sentence coherence.
The results provide quantitative support that our model can generate grammartically correct and lexically coherent and diverse story texts.

\begin{table}[]
\caption{Automated evaluation of human-written and machine-generated stories using lexical metrics.}
\label{table.lexical_eval}
\resizebox{0.9\linewidth}{!}{%
\begin{tabular}{lllll}
\hline
 &
  \textbf{Grammar} &
  \textbf{\begin{tabular}[c]{@{}l@{}}Type-token\\ ratio\end{tabular}} &
  \textbf{\begin{tabular}[c]{@{}l@{}}Trigram\\ repetition\end{tabular}} &
  \textbf{\begin{tabular}[c]{@{}l@{}}Sentence\\ coherence\end{tabular}} \\ \hline
Human &
  1.00 &
  0.75 &
  0.01 &
  0.42 \\
Machine &
  1.00 &
  0.77 &
  0.01 &
  0.43 \\ \hline
\end{tabular}%
}
\end{table}

\begin{table}[]
\caption{Average human ratings of machine-generated and human-written stories. (*: p < .05 using Wilcoxon Signed-rank test)}
\label{table:human_rating}
\resizebox{\linewidth}{!}{%
\begin{tabular}{lllll}
\hline
        & \textbf{Coherence *} & \textbf{Relevance *} & \textbf{Interestingness} & \textbf{Overall} \\ \hline
Human   & 4.53               & 4.58               & 4.08                     & 4.26                     \\
Machine & 3.92               & 4.33               & 3.97                     & 4.01                     \\ \hline
\end{tabular}%
}
\end{table}



\subsubsection{Human Evaluation}
We invite eight PhD students (four females and four males, mean age: 25.50 (SD = 2.07)) with English paper publications to rate their perceived quality of these 40 stories in random order. 
According to previous work~\cite{yao2019plan, GoldfarbTarrant2019PlanWA}, we consider: \textbf{coherence} (\textit{The story is logically consistent and coherent}), \textbf{relevance} (\textit{The story is relevant to the title}), 
\textbf{interestingness} (\textit{The story is interesting}), and \textbf{overall quality} (\textit{Overall, it is a good story}). 
Each aspect is rated on a standard five-point Likert Scale (1  for ``Strongly disagree'' and 5 for ``Strongly agree''). 
As shown in \autoref{table:human_rating}, 
the machine-generated stories achieve comparable performance with the human version regarding overall quality and interestingness. 
The human-written stories are considered significantly more coherent and relevant using the Wilcoxon Signed-rank test.
Nevertheless, the machine-generated stories have average scores of around four points in terms of coherence and relevance. 
Therefore, we consider that our system could produce adequate story context given a set of target words for vocabulary learning. 

\subsection{Exploring Vocabulary Learning Activities with Story Generation Models}
After validating the feasibility of our model for generating meaningful context that covers a set of CET-4 English words, we explore possible vocabulary learning activities that the model can support. 
We conduct semi-structured interviews with three English teachers (E1-3, age: 27 - 28) and five university students (S1-5, age: 21 - 29) in China. 
E1 has two years of experience in teaching IELTS and half-a-year experience in teaching English in a higher vocational college.
E2 has spent five years in high-school English teaching, and E3 has taught high-school students mainly about TOEFL writings for three years in an educational institution.
S1-5 are well-experienced in using different English vocabulary learning software for Chinese (\eg Liulishuo, Baicizhan, Shanbay). 
\subsubsection{Procedure} Each interview starts with participants' practices (whether, why, and how) of story-based activities for teaching or learning vocabulary. 
Then, we show participants a web interface that allows users to input target English words and generate stories with those words based on our model. 
We prepare example CET-4 word sets, each with the top-five topic-relevant words (\eg cable, complain, library, instruction, unfortunate) \footnote{We use sentence-bert \cite{reimers2019sentence} to encode the words into vectors and rank them based on their cosine similarities with the vector of the encoded title.} under our specified titles (\eg the internet) and the generated stories in the interface. 
We invite our participants to check the generated stories and have a trial using their specified words. 
During this process, we encourage them to brainstorm the vocabulary learning activities that our story generation model can support. 
Each interview lasts for about 30 minutes with about USD \$3.5 for compensation.

\subsubsection{Results}\label{sec:interview_teachers} We transcribe the audio data into texts and group them into themes following the interview structure. Both groups of interviewees confirmed that learning English vocabulary via stories is a common and effective practice. 
For example, E1 mentioned that he usually asks students to first write sentences and then create a short story with newly learned words, which helps them master the usage of words. 
Three student participants regularly read English books and articles, which expands their vocabulary. 
In general, all participants agreed that our generative stories are suitable materials for learning target words. 
For instance, E3 tried the story generation model using ``health'' as title and ``tobacco, alcohol, abuse, dominate, harmful'' as target words. 
These words come from an article of her high-school text book. 
\textit{``I like the generated story. It is generally coherent, and it is simpler than the one in the text book. My students would like it for vocabulary learning as they do not need to pay too much attentions on the long sentences''} (E3). 
Nevertheless, our three teachers pointed out that the generated stories lack explicit logic transition words like ``nevertheless'' and ``for example'', which can further improve the stories' coherence. 
This is probably due to the lack of these words in our training dataset ROCStory corpus. 


Our interviewees actively provide ideas for leveraging our generative model to support vocabulary learning. 
Together with the insights from pedagogical literature (\eg those in \autoref{subsection_2.1}), we summarize three supported vocabulary learning activities.

\textbf{Story reading}: learners can read the generated story to understand the meanings of the target words (E1-3, S1-5). This is also a typical meaning-focused input activity suggested by language educators \cite{nation2007four}. 

\textbf{Cloze test}: learners can do a cloze test that fills blanks of the generated story using target words to strengthen their understandings (E2, S2, S3). \textit{``Cloze test is a common vocabulary learning strategy in the textbook. I feel that it would be helpful to customize the generated stories into cloze tests for students''} (E2). Cloze test can be viewed as a language-focused learning activity with a focus on the usage of target words \cite{nation2007four}. 

\textbf{Turn-taking writing}: learners can take turns with the generative models to co-write a story using target words (E1, E3). 
This practice combines the meaning-focused output and fluency development learning activities, as it requires learners to use the learned words productively and fluently \cite{nation2007four}. 
\textit{``It can generate a sentence using a target word as a start, and users write down the second sentence using another word. With such interaction, students can learn how to use the words in a successive manner''} (E1). 
\textit{``The system can act as one student to do a turn-by-turn, co-writing practice''} (E3). 
The system can provide in-situ guidance and feedback on users' input in the writing process \cite{nation2007four}, \eg
\textit{``are target words used correctly''} (E3) and \textit{``is the story coherent and correct in grammar?''} (E1).

\section{Phase 2: \name{} System Implementation and Refinement}
After validating the feasibility of our story generation model and identifying promising ways to apply it, we present our second-phase design process about how we implement \name{} and refine it with feedback from learners, teachers, and HCI researchers. 

\begin{figure*}
    \includegraphics[width=\textwidth]{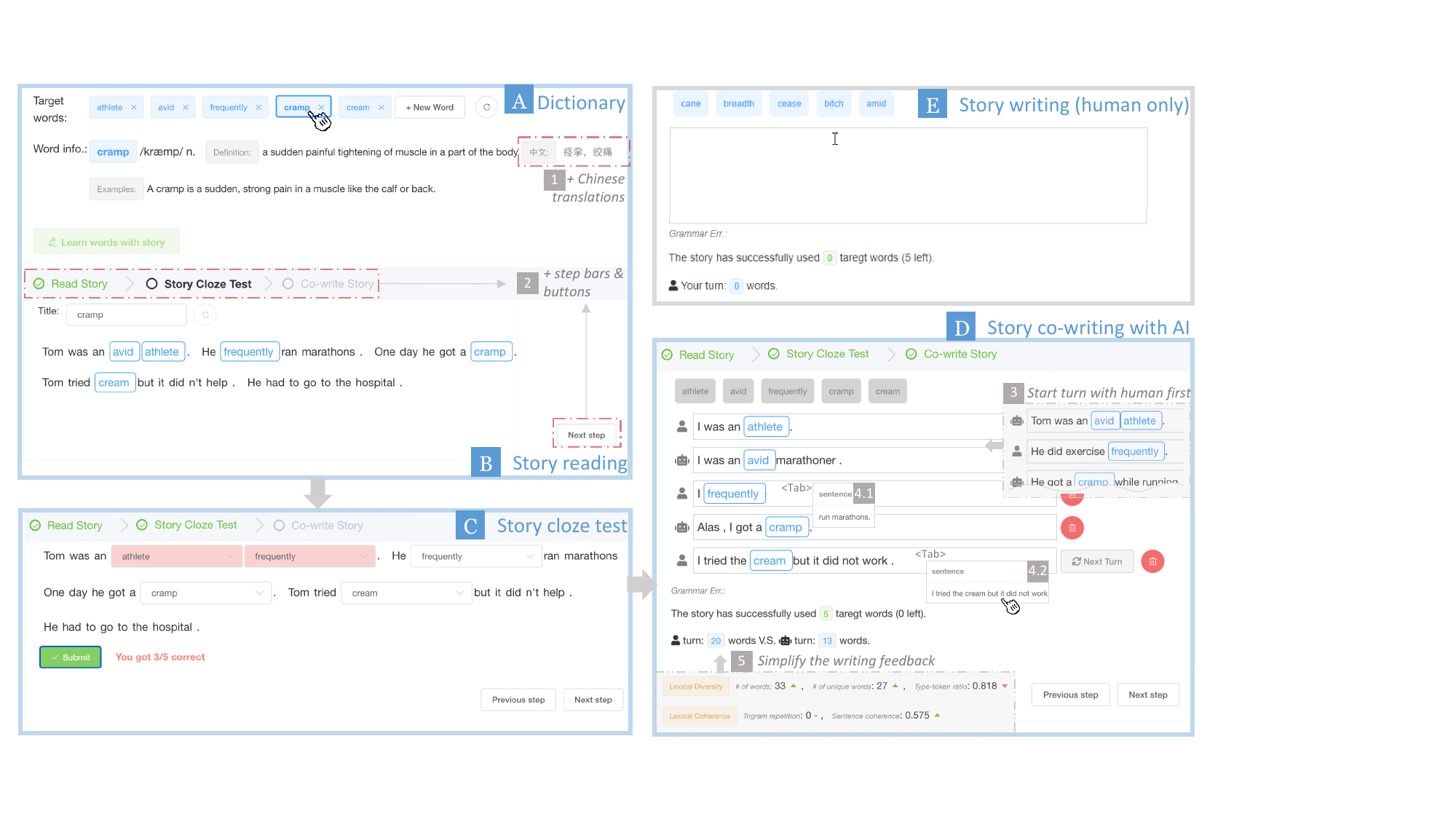}
    \caption{
    The interface designs of \name{}. 
    (A) Users can specify target words and check their meanings.
    (B) Story reading: users can read a machined-generated story that contains target words. 
    (C) Cloze test: users can conduct a cloze test by using target words on the generated story. 
    (D) Story writing: users can take turns with \name{} to write a new story using target words. 
    (E) The interface for story writing without adaptive support in the Storyfier-sen baseline (\autoref{sec.eval}). 
    Note that in the first \name{} prototype, the three modes (B-D) are separated. 
    In the refined \name{}, we unify them into one flow and improve the system designs (1-5) based on feedback from learners and experts. 
    }
    \label{fig:system}
    \Description{Figure 3 shows the interface designs of Storyfier. Five interfaces are labelled (A) Dictionary, (B) Story reading, (C) Story cloze test, (D) Story co-writing with AI, and (E) Story writing (human only). The content labelled 1, 2, 3, 4, and 5 indicates the refinement of Storyfier, as described in the main text. The click on the “Next step” button leads to the transition between the interfaces, i.e., from B to C, from C to D, or from C to E. }
\end{figure*}

\subsection{\xb{First \name{} Prototype with Three Modes}}

Based on the interview findings, we design and implement three modes of user interfaces to facilitate vocabulary learning via story reading (\autoref{fig:system}B), cloze test (C), and turn-taking writing (D). 

\subsubsection{\xb{Interface Designs and User Workflow}}
All three modes share the following two features (\autoref{fig:system}A). 
[\textbf{Target words setting}]
Users can manually add new words (``+'') or delete them (``x'') as they wish. They can also click the \faRepeat{} button to get a randomly sampled target word set.
[\textbf{Dictionary lookup}]
Users can click each target word to inspect its definition, part of speech, phonetic symbol, 
and usage example. 
The click on \faPencil{} 
will lead to three vocabulary learning activities supported in the following interface variants. 

\textbf{Story reading} mode.
This interface (\autoref{fig:system}B) presents the AI-generated story with the target words highlighted in blue, which could help users quickly inspect their contextual use. In addition, users can conduct minor edits (\eg, revise words) of the sentences to refine the story if they wish. 

\textbf{Cloze test} mode.
This interface (\autoref{fig:system}C) replaces the target words in the generated stories with blanks. Users are required to make contextual inferences about the missing words and choose the proper ones to fill the blanks. After they submit the results,  \name{} will check the correctness and highlight the misused words (if any) in red. Users can iteratively fix the errors if they wish. 

\textbf{Turn-taking writing} mode.
This interface (\autoref{fig:system}D) encourages users to write a story with AI using the target words sentence by sentence.
During the writing process, 
users can gain an overview of the used (gray) and unused (blue) target words at the top in \autoref{fig:system}D. 
The used target words are highlighted in the corresponding sentences. 
To provide adaptive feedback to learners, in each turn, 
the system will check and alert the grammar issues of the written text using LanguageTool API\footnote{\url{https://languagetool.org/http-api/}}.
Meanwhile, \name{} provides writing feedback on the story sentences at the bottom regarding grammar errors, lexical diversity, and lexical coherence (in \autoref{fig:system}-5). 
Red and green triangles indicate a decrease or increase in scores of all current story sentences compared to the one in the previous turn.
Users can write and refine the story with \name{} until all the target words are used. 



\subsubsection{\xb{Controllable story generation model that supports turn-taking writing}}
\label{subsec.control_generation}
To support the turn-taking writing activity where \name{} needs to produce varied-length text spans given previous story sentences and target words, we further build a story-infilling model (\autoref{fig:algorithm}B-2).
We formulate the training objective as a span prediction task and adopt a prompt-based approach to finetune the T5 model. 
Given a ROC story, we derive its story title, target words, and story sentences as prompts (described in \autoref{subsec.story_generation}).
Meanwhile, for each story sentence, 
we randomly mask varied-length of text spans of this sentence,
following prior work \cite{Donahue2020EnablingLM}. 
Then, we train the model to predict the masked spans of the current sentence based on the prompts. 
We use a cross-entropy loss and finetune the pretrained T5-large model provided by Huggingface on our mask prediction task using adam optimization algorithm. We train 10 epochs, and the training loss is 1.0701.

With this finetuned model, \name{} can write the next sentence using target words following the users' written ones. 
Furthermore, in our refined \name{} presented below (\autoref{subsec.refined_sys}), it can also help users revise an existing sentence or complete the unfinished one via text infilling. 

\subsection{Testing \name{} Prototype}
To seek feedback on the 1st \name{} prototype, we conduct a usability test with ESL learners and two workshops with English teachers and HCI researchers.  

\subsubsection{Usability Test with 12 ESL Learners.} 
\label{subsec.usability_test}
\label{section_4.2.1}
To probe the user experience and perceived usefulness of the three activities supported by \name{}, we conduct a within-subjects usability test with 12 junior undergraduate students ($6$ females, $6$ males, mean age: $19.5$ (SD = $0.52$)) in a university in China. 
The baseline condition does not have the generated story contexts but provides a \textit{dictionary} function that shows the meanings, synonyms/antonyms, and usage examples for each target word (\autoref{fig:system}A1).  
All participants have passed the national English exam CET-4, with an average score 560.50 (SD = 37.75) \footnote{425/710 points are considered passed for CET-4.}. 
We do not aim to evaluate \name{}'s effectiveness but seek to improve it with quick user feedback at this stage. 
These participants can provide us with valuable feedback as they have fresh CET-4 vocabulary learning experience. 

[\textit{Procedure}]
Participants use their own computers to remotely conduct the study following the instructions. 
They experience the four experiment conditions (i.e., \textit{Dictionary, Read, Cloze, Turn-taking Write}) one by one in a Latin-Scale counterbalanced order. 
In each condition, they learn two prepared word sets, each with five CET-4 words sampled based on topic relevance. 
After each condition, participants rate their perceived usefulness, easiness to use, and intention to use \cite{AL_adaptive_learning10.1145/3313831.3376732, Technology_acceptance_model:doi:10.1111/j.1540-5915.2008.00192.x} of each interface in a 7-points Likert scale; 7 for a strong agreement. 
In the end, we ask for their comments and suggestions on \name{}.  
They receive about USD \$9.5 for around 50 minutes spent in the study.

[\textit{\textbf{Results}}] We use repeated measured ANOVA test (\textit{Dictionary} vs. \textit{Read} vs. \textit{Cloze} vs. \textit{Turn-taking Write}) to evaluate the user experience of \name{}'s three modes. 
There is a significant difference in perceived usefulness of the four activities; $F(3, 33) = 3.83, p < 0.05, \eta^2 = 0.26$. 
Specifically, they feel that the \textit{Read} ($M = 4.94, SD = 1.15, p < 0.05$) system is significantly more useful than the \textit{Dictionary} one ($M = 3.56, SD = 1.45$). 
Participants feel that the \textit{Read} ($M = 4.83, SD = 1.12, p < 0.01$) system is significantly easier to use than the \textit{Turn-taking} one ($M = 3.15, SD = 0.85$); $F(3, 33) = 9.54, p < 0.001, \eta^2 = 0.46$; Bonferroni post-hoc test. 
Besides, the \textit{Cloze} ($M = 4.27, SD = 1.13, p < 0.05$) system is deemed significantly easier to use than the \textit{Turn-taking} one. 
Lastly, participants have significantly higher intentions to use the \textit{Read} ($M = 4.71, SD = 0.33$) system for their vocabulary learning in the future, compared to the \textit{Dictionary} ($M = 2.96, SD = 1.63, p < 0.01$) and \textit{Turn-taking} systems ($M = 3.13, SD = 1.40, p < 0.05$); $F(3, 33) = 5.80, p < 0.01, \eta^2 = 0.35$. 
In summary, ESL learners found that the three learning activities supported by \name{} are more useful than the baseline without story context. 
However, the \name{}'s turn-taking writing mode should be further improved. 
For example, two students indicated that sometimes they found it difficult to use words to write the next story sentence in this activity. 

\subsubsection{Co-Design Workshops with English Teachers and HCI Researchers.} 
\label{subsec.co-design_workshop}
Apart from the feedback from ESL learners, we conduct two co-design workshops to seek experts' feedback. 
The two workshops share a similar procedure but have a different focus. 
One is with the same three English teachers (E1-3) in Phase 1 and focuses on refining the vocabulary learning activities in \name{}. 
The other is with four HCI researchers (H1-4, all males, age: 25-27) and mainly works on the interface and interaction design of \name{}. 
All HCI researchers have experience in developing intelligent systems and have papers published in top venues like CHI and VIS.
Each workshop starts with a warm-up activity in which participants share their experience of story-based vocabulary teaching or learning. 
Then, we show our \name{} prototype to them, invite them to have a trial, and ask them to give comments on the system. 
Next, we organize a brainstorming session to discuss how to leverage the three learning activities of \name{} for effective vocabulary learning support and how to improve the interaction and interface design.
Each workshop lasts about one hour, and participants receive about USD \$17 as compensation. 
We present their suggestions on \name{} together with its refinement in the next subsection.

\subsection{\xb{Refined \name{} System}}\label{subsec.refined_sys}

Based on the collected feedback on the first \name{} prototype, we refine its workflow and features (\autoref{fig:system}). 

\textbf{Workflow}. We unify the three separate learning activities into one workflow using step bars and next-step buttons (in \autoref{fig:system}-2) to guide learners to read the story, do a cloze test, and write a new story. 
Our English teachers agree that all three learning activities would be generally helpful, but there could be a flow that chains these activities to maximize their values. 
They suggest that \textit{reading} should be the first activity to help comprehend the target words. 
The cloze test should come next to strengthen their understanding, and the co-writing practice should be the last activity.
\textit{``Cloze test is a controlled practice, and co-writing is a free one''} (E3). 
\pzh{
To chain the three learning activities into a flow, S3 proposes to use a chatbot to guide users through the learning process, which could be engaging. 
This is similar to the chatbot interaction in StoryBuddy \cite{zhang2022storybuddy}. 
However, the other three HCI researchers are concerned that it might distract users' attention from vocabulary learning to interaction with the chatbot. 
S1 suggests that we can use clear widgets (e.g., the right arrow and ``Next Turn'' buttons) to order the flow of the three activities. 
}

\textbf{Features.} 
First, we add the main Chinese meaning of each target word in the dictionary (\autoref{fig:system}-1) as suggested by E2.  
Second, we modify the turn-taking order by encouraging users write the first sentence of the story (\autoref{fig:system}-3), as suggested by E3 that we should encourage learners to spend effort first. 
Third, we add an inline sentence suggestion function that can infill a generated next sentence using target words (\autoref{fig:system}-4.1 and -4.2), to address learners' difficulties in story writing activity found in the usability test. 
This function can be triggered in real-time by the ``tab'' key on users' demands, as suggested by the HCI researcher H4.
Third, we remove the technical metrics about sentence quality (\autoref{fig:system}-5), as suggested by E1-3 that they are complicated for learners and not focused on the target words' usage. 
Fourth, we add the number of used target words and the number of words written by human/machine as writing feedback because it could encourage learners to write more. 




\section{Experiment} \label{sec.eval}
To explore how would \name{} impact the users' vocabulary learning outcome and experience, we conduct an experiment with 28 ESL (English-as-the-Second-Language) Chinese students. 
We adopt a 2 (with vs. without AI features) x 2 (read-only vs. read-cloze-write activities) within-subjects design. 
The first one -- \textbf{AI factor} -- aims to study the impacts of \name{}'s AI-generated story and adaptive writing support. 
\pzh{We note the conditions with AI features with ``-AI'' and those without AI features with ``-sen''.}
The second one -- \textbf{activity factor} -- identifies the value of additional cloze test and writing activities to the reading activity that previous vocabulary learning support systems focus on. 
\pzh{We note the conditions with the read-only activity with ``Read-'' and those with read-cloze-write activities with ``Storyfier''.}
The four conditions are: 

\pzh{\textbf{Read-only}}
\begin{itemize}
    \item \textbf{Read-sen} interface only provides dictionary features with an example existing sentence for each target word (\autoref{fig:system}A); 
    \item \textbf{Read-AI} interface additionally provides a generated story that covers target words (\autoref{fig:system}A + B); 
\end{itemize}

\pzh{\textbf{Read-cloze-write}}
\begin{itemize}
    \item \textbf{Storyfier-sen} interface offers example sentences for target words, a cloze test on these sentences, and a writing exercise without AI's intervention (\autoref{fig:system}A + B + D, but the stories in B and C are replaced by the example sentences of target words); 

    \item \textbf{Storyfier-AI} interface contains all features of \name{} (\autoref{fig:system}A + B + C). 
\end{itemize}

The Read-sen and Storyfier-sen interfaces simulate how individuals traditionally use existing materials to learn any target word set without adaptive support, which can help us evaluate the impact of \name{}'s story generation model. 

Our research questions are: 

\textbf{RQ1.} How would \name{} affect vocabulary learning outcome? 

\textbf{RQ2.} How would \name{} affect the learning experience? 

\textbf{RQ3.} What are user perceptions towards \name{}?

\subsection{Participants}
We recruit 28 second-year undergraduate students (P1-28, $24$ females, $3$ males, $1$ prefer not to tell, mean age: $20.04$ (SD = $0.69$)) from a course in a college in mainland China. 
They are typical ESL learners who major in Business English. 
The course nature leads to the gender and major unbalance of our participants, which we will discuss in the Limitations subsection. 
Twelve of them have not passed the national English exam CET-4 in China, and the rest have passed it with an average score $493.8 / 710$ ($SD = 35.8$) \footnote{425/710 points are considered passed for CET-4.}.
Their self-assessed English vocabulary proficiency score is $4.39$ (SD = $0.63$; $1$ - not proficient at all, $7$ - very proficient). 

\subsection{Procedure and Tasks}
\begin{figure}
\includegraphics[width=0.45\textwidth]{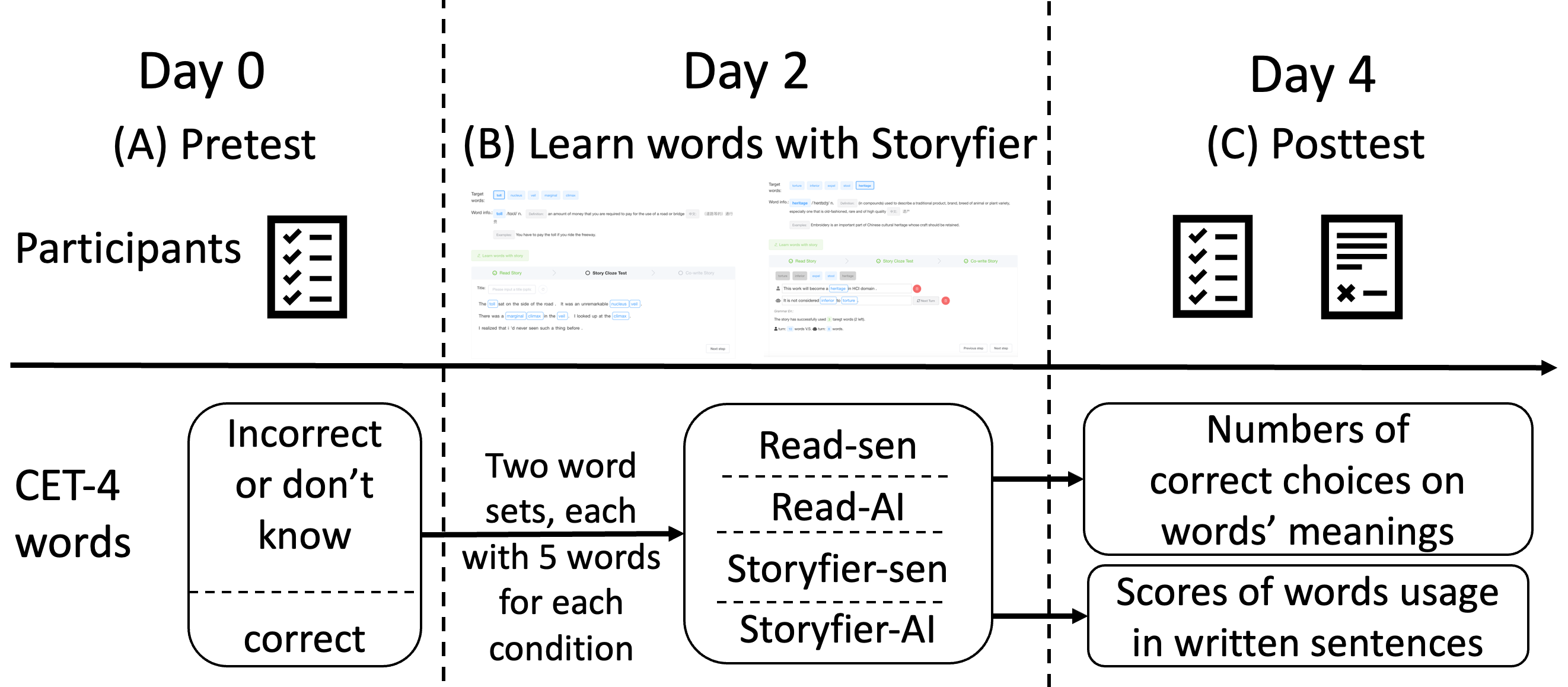}
    \caption{Procedure of the experiment. (A) Participants first took a pretest, and the words they did not know were target words. (B) On the experiment day, they used the four interfaces of Storyfier for vocabulary learning. (C) Two days later, they took the posttest on words' meanings and usage.}
    \label{fig:procedure}
    \Description{Figure 4 shows the procedure of the experiment. As described in the main text, participants need to take a Pretest on Day 0, Learn words with Storyfier on Day 2, and take a Posttest on Day 4. }
\end{figure}

We conduct the experiment remotely. 
In a similar manner as \cite{VocabEncounter}, the procedure of our experiment consists of three stages (\autoref{fig:procedure}). 
First, after collecting the background information with consent, we ask each participant to take a pretest to identify the CET-4 words that they did not know. 
In the pretest, the participant needs to choose one of the five options, including four meanings written in Chinese and one indicating ``I do not know this word'', for each CET-4 word.
We invite a postgraduate to prepare 170 CET-4 words that are not easy (\eg excluding words like ``easy'' and ``feel'') from the English learning app \textit{Baicizhan} and only include the intended participants who answer incorrectly or indicate lack of knowledge on at least 40 words. 
For each participant, we randomly select 40 of the identified unknown words and divide them into eight sets, each with five target words. 
After the pretest, we also have participants read the instructions of the learning tasks and the four interfaces. 
We inform them not to learn the words that appear in the pretest prior to learning sessions. 

Then, on the experiment day, participants log in to their learning sessions via their unique IDs. 
Participants are asked to learn two word sets with each \name{} interface. 
We counterbalance the order of the four interfaces using Latine Square. 
After learning two word sets with an interface, participants rate their engagement and enjoyment in the learning process, perceived learning performance, and perceptions of the system in a questionnaire. 
Upon completion of four tasks, we further ask for their preferences on the interfaces, comments on the generated stories and AI's writing assistance, and suggestions for improving \name{}.

Next, two days after the experiment day, we ask the participants to take a posttest, which has a similar format as the pretest but only presents the 40 words they met in the learning sessions. 
In the posttest, participants also need to write a sentence for each target word if they do not choose ``I do not know this word''. 
They can write ``nothing'' if they feel hard to write the sentence. 
Each participant spends about 1.5 hours in total on the full procedure and gets around USD \$12 as compensation. 

\subsection{Measurements}
\textbf{RQ1. Learning Outcome.} We measure participants' retention of target words' meanings via the number of correct answers to the multiple-choice questions in the posttest. To capture how well they learn the usage of target words, we invite one English teacher (E1 in our workshop) to rate the grammar correctness (\eg tense and part of speech) and context appropriateness of the target word in each written sentence in the posttest using a three-point scale; 0 - not correct, 1 - partially correct, 2 - correct. 
For each participant in each system interface, we calculate i) the numbers (range: 0 - 10) of sentences that use target words correctly in terms of both grammar and context and ii) the total score (0 - 40)  of sentences \footnote{In each interface, participants learn ten words and write at most ten sentences in posttest. The maximum score for each sentence is $2 + 2 = 4$.}. 

\textbf{RQ2. Experience.} We measure users' engagement and enjoyment during the learning process with each system interface (\textit{``I was absorbed in using this interface to learn vocabulary''}  and \textit{``It is enjoyable to learn vocabulary with this interface''} \cite{arguetutor,engagement}). 
Besides, we measure the perceived task workload of learning sessions using items adapted from NASA Task Load Index \cite{hart2006nasa} (\eg ``I have to work hard to accomplish the writing activity.''). 
Apart from the questionnaire data, we also log the i) task completion time of learning two word sets with each interface, as well as ii) the amount of time spent in reading, cloze-test, and writing activities and iii) written stories in Storyfier-sen and Storyfier-AI interfaces. 

\textbf{RQ3. Perceptions towards \name{}.} We adapt the technology acceptance model \cite{AL_adaptive_learning10.1145/3313831.3376732, Technology_acceptance_model:doi:10.1111/j.1540-5915.2008.00192.x} to the perceived usefulness (four items, \eg \textit{``The use of this interface enables me to learn the vocabulary more efficiently''}; Cronbach's $\alpha = 0.944$), easiness to use (four items, \eg \textit{``I would find this interface to be flexible to use''}; $\alpha = 0.786$), and intention to use (two items, \eg \textit{``If this interface is available there to help me learn vocabulary, I would use it''}; $\alpha = 0.966$) of each system. 
We average the ratings of multiple questions as the final score for each aspect.
All statements in the questionnaire are rated on a standard 7-point Likert Scale, with 7 for a strong agreement.

\section{Analyses and Results}
For the rated items, we first check whether the order of the four experienced interfaces affects our results via a set of mixed ANOVA tests (order as between-subjects, interfaces as within-subjects) on each rating. Neither the main effect of the order nor its interaction effect with the system interface is significant. 
Hence, except those with additional notations (\eg one-way ANOVA),  the statistic tests in this section are two-way repeated measured ANOVAs. 
For each ANOVA, the assumption of equal variance holds according to Macuchly's test of sphericity 
\cite{ANOVA}. 
For the participants' comments on \name{}, two authors conduct an inductive thematic analysis \cite{thematic_analysis}. 
They first independently assign codes to the text data and then discuss the codes for several rounds. 
After that, they group the codes into categories, which are incorporated into the results below. 


\subsection{RQ1: Impact on Learning Outcome} 
\label{subsection:6.1}
\begin{figure}[t]    \includegraphics[width=0.48\textwidth]{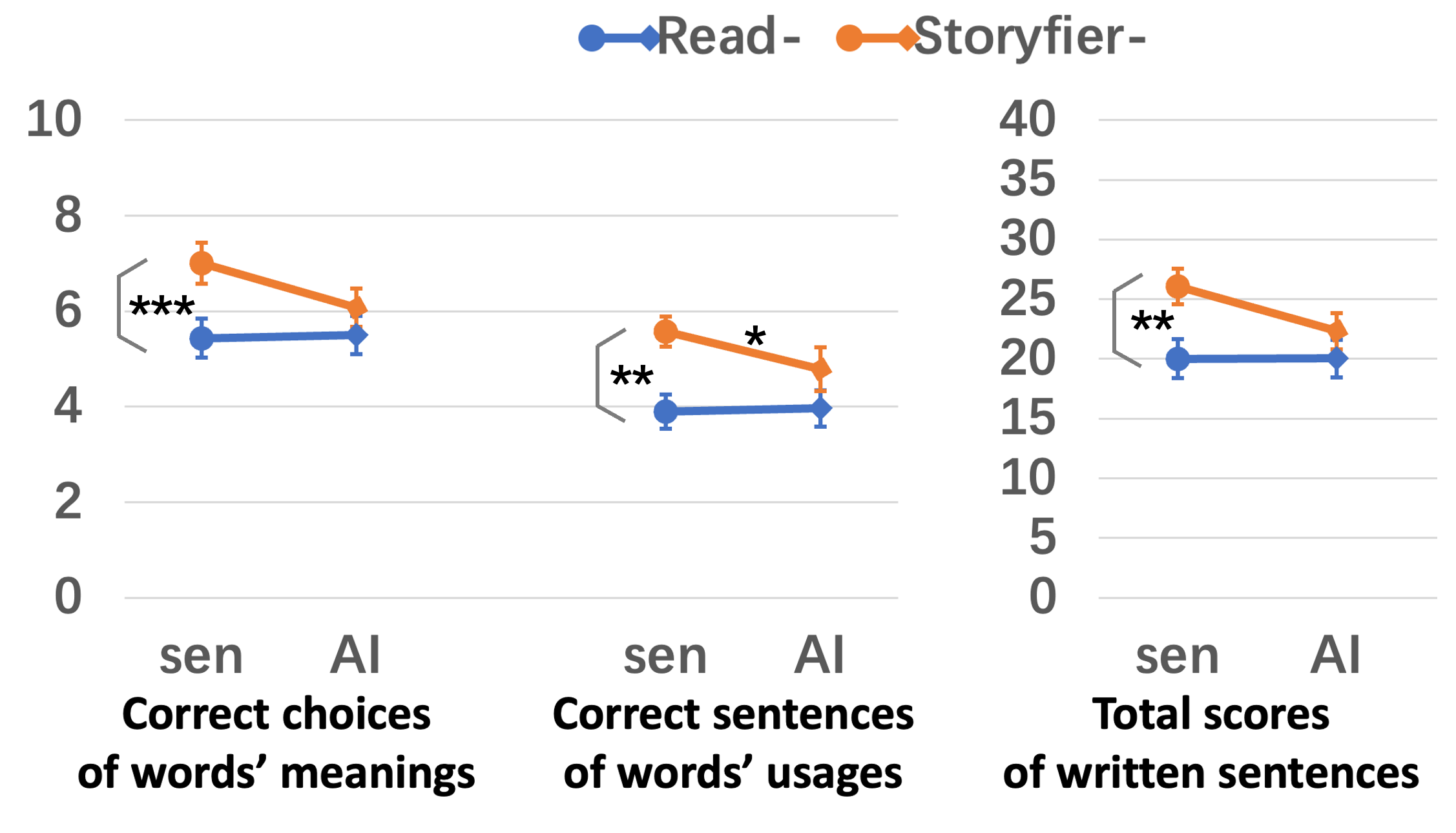}
    \caption{RQ1 results regarding numbers of correct choices on target words' meanings, numbers of sentences that correctly use target words, and total scores of the written sentences in each condition. ***: $p < 0.001$, **: $p < 0.01$, *: $p < 0.05$.}
    \label{fig:RQ1}
    \Description{Figure 5 shows a line graph that consists of three parts, each with two lines that represent Read and Storyfier interfaces, as described in the main text. Each line shows the comparison of the metric between the interfaces without (noted as sen) or with AI features. The Y axis in the first two parts represents a number ranging from 0 to 10. The Y axis in the last part represents a total score ranging from 0 to 40. The lines of Storyfier interfaces are significantly higher in Y axis than the lines of Read interfaces. }
\end{figure}

\autoref{fig:RQ1} shows the results regarding learning outcomes. 
\subsubsection{Retention of target words' meanings}
Our results indicate that neither the AI factor nor its interaction with the activity factor significantly affects the retention of target words in four conditions. 
However, the Storyfier-sen interface results in a better retention performance ($M = 7.00, SD = 2.16$) than the Storyfier-AI interface ($M = 6.07, SD = 2.09$); $p = 0.049$, one-way repeated-measures ANOVA. 
Besides, participants perform significantly better in target words' retention in the read-cloze-test \pzh{(\ie Storyfier-sen and Storyfier-AI)} conditions ($M = 6.54, SD = 2.19$) than that in the read-only \pzh{(\ie Read-sen and Read-AI)} conditions ($M = 5.46, SD = 2.16$); $F = 9.605, p = 0.004$. 

\subsubsection{Target words' usage in the sentences} 
Neither the AI factor nor its interaction with the activity factor has a significant impact on i) the number of sentences that correctly use target words and ii) the total scores of written sentences in four conditions. 
However, when comparing the means between the Storyfier-sen and Storyfier-AI interfaces, we \pzh{observe that}
in the read-cloze-write learning sessions, Storyfier's AI features could reduce learning gains on target words' usage. 
As for the activity factor, our results show that participants with the read-cloze-write interfaces ($M = 24.20, SD = 8.66$) perform significantly better in word usage in their written sentences than the cases with the read-only interfaces ($M = 20.02, SD = 7.92$); \eg for ii) total scores, $F = 12.721, p = 0.001$. 

\pzh{In all, we find that \name{}'s AI features reduce learning gains on the retention of target words' meanings in the read-cloze-write vocabulary learning sessions. 
Its supported additional cloze-test and writing practices improve learning gains on target words' meanings and usage compared to learning via reading-only activities.}

\subsection{RQ2: Impact on Learning Experience}
\begin{figure*}
\includegraphics[width=\textwidth]{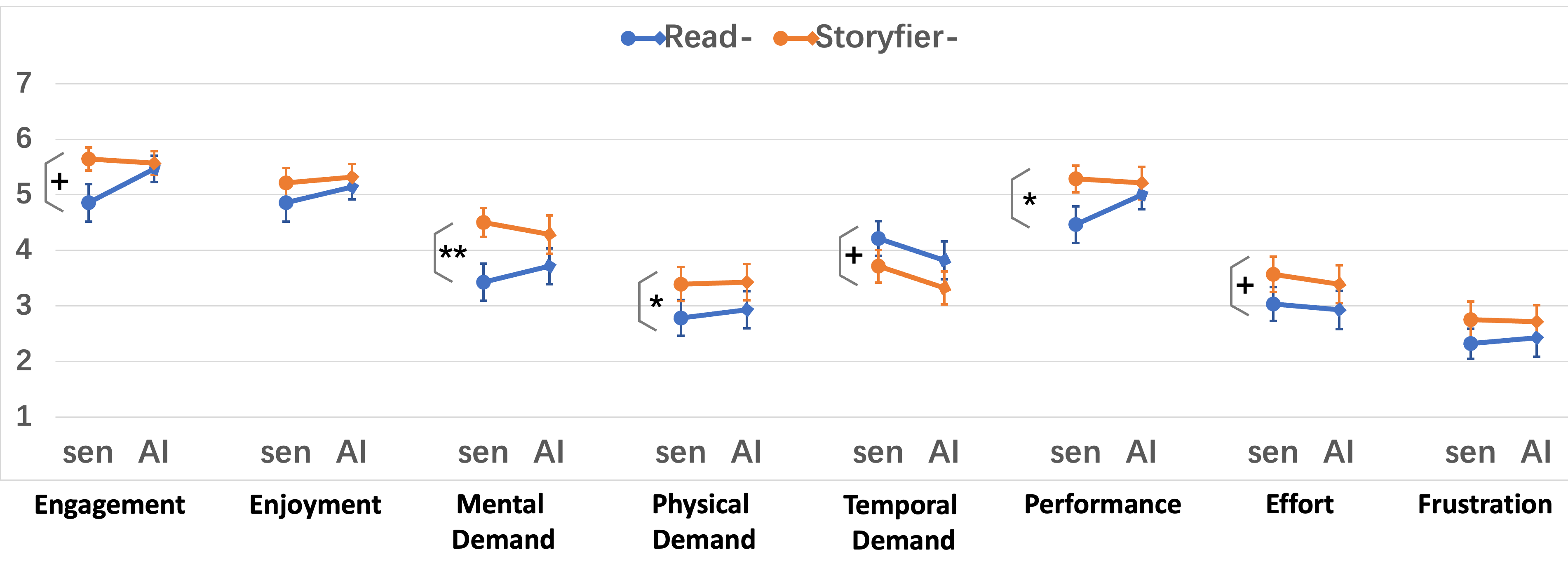}
    \caption{RQ2 results regarding perceived engagement, enjoyment, and workload in vocabulary learning sessions with Read-sen, Read-AI, Storyfier-sen, and Storyfier-AI interfaces. **: $p < 0.01$, *: $p < 0.05$, +: $p < 0.1$.}
    \label{fig:RQ2}
    \Description{Figure 6 shows a line graph that consists of eight parts, each with two lines that represent Read and Storyfier interfaces, as described in the main text. Each line shows the comparison of the metric between the interfaces without (noted as sen) or with AI features. The Y axis represents the Likert scale ranging from 1 to 7. The lines of Storyfier interfaces are generally higher in Y axis than the lines of Read interfaces, except the lines in the Temporal Demand part. }
\end{figure*}
\subsubsection{Engagement, enjoyment, and workload}
As shown in \autoref{fig:RQ2} \footnote{The full statistics are attached in the supplementary materials.}, 
neither the AI factor nor its interaction with activity factor significantly affects users' perceptions on their learning experience. 
When digging into each measured item in each condition, we have several interesting observations: 
a) in read-only sessions, participants with AI-generated stories could feel more engaged and enjoyed than the cases without these stories; 
b) \name{}'s AI features could increase mental demand and perceived performance in read-only learning sessions but decrease the ratings on these measures in read-cloze-write sessions; 
c) \name{}'s AI features could reduce temporal demand, \ie how rushed is the pace of the task, and perceived spent effort in the vocabulary learning tasks. 
As for the activity factor, we found significant differences regarding the perceived mental demand ($p = 0.002$), physical demand ($p = 0.029$), and perceived performance ($p = 0.025$). 
\pzh{Specifically,} Storyfier's supported additional cloze-test and writing practices increase mental and physical demand and perceived performance compared to learning via reading-only activities.
We also \pzh{observe} that these additional activities could increase engagement and spent effort in the vocabulary learning tasks. 

\subsubsection{Task completion time and written stories} 
i) On average, participants spent $89.22 (SD = 105.25)$ / $226.52 (150.66)$ / $805.00 (490.55)$  / $806.61 (368.19)$ seconds in the learning session with Read-sen, Read-AI, Storyfier-sen, or Storyfier-AI interface. 
\pzh{This indicates that} in the read-only vocabulary learning sessions, participants spent significantly more time when they were presented with AI-generated stories than when they were not ($p < 0.001$). However, as shown in \autoref{fig:RQ1}, the learning gains do not increase accordingly.
ii) When digging into the average amount of time spent in each learning activity for each word set in Storyfier-sen and -AI interfaces, we have $49.48$ vs. $81.56$ (read, $p = 0.004$), $30.76$ vs. $43.84$ (cloze, $p = 0.004$), and $263.46$ vs. $208.71$ (write, $p = 0.09$) seconds \footnote{The total time does not match task completion time as it does not include time spent on checking each word's meaning.}. 
\pzh{This shows that} compared to the sessions with Storyfier-sen, participants with Storyfier-AI spent significantly more time in reading and cloze-test activities but less time in writing activities.

\subsection{RQ3: Perceptions on \name{}}
\begin{figure}
\includegraphics[width=0.48\textwidth]{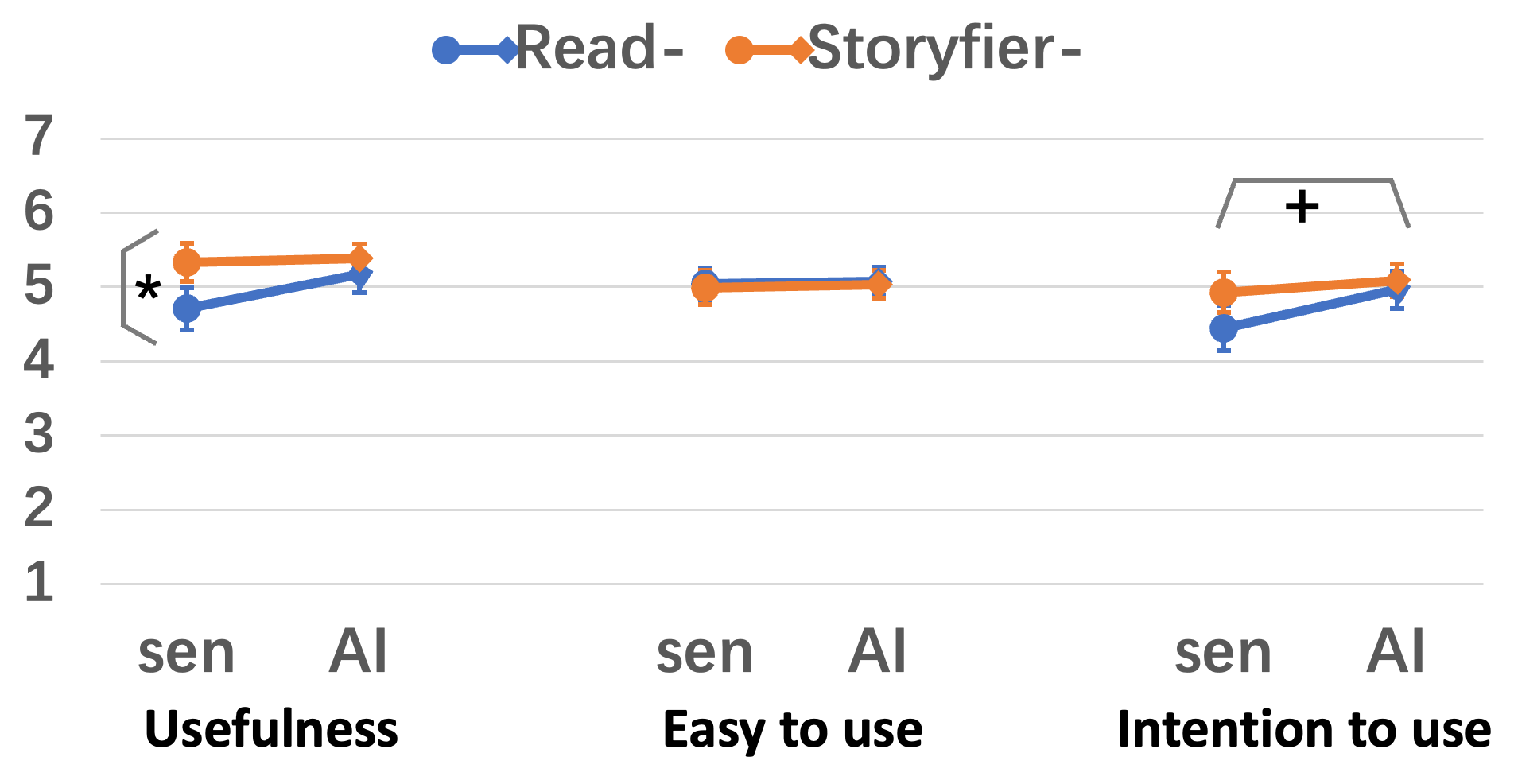}
    \caption{RQ3 results regarding user perceptions with each interface. *: $p < 0.05$, +: $p < 0.1$.}
    \label{fig:RQ3}
    \Description{Figure 7 shows a line graph that consists of three parts, each with two lines that represent Read and Storyfier interfaces, as described in the main text. Each line shows the comparison of the metric between the interfaces without (noted as sen) or with AI features. The Y axis represents the Likert scale ranging from 1 to 7. The lines of Storyfier interfaces are generally higher in Y axis than the lines of Read interfaces, except the lines in the Easy to use part.  }
\end{figure}

\autoref{fig:RQ3} depicts users perceptions of each \name{} interface. 
\subsubsection{Quantitative items}
In read-only sessions, there is a trend on the improved perceived usefulness and intention to use of the interfaces with AI-generative stories over those without the stories. 
This implies \pzh{that in} read-only learning sessions, participants would find Storyfier more useful and have a higher intention to use it  if it provides AI-generated stories.
As for the activity factor, \pzh{ participants feel that} \name{} is significantly more useful ($p = 0.035$) if it supports cloze test and writing practices in addition to the reading activities.
There are no significant differences in the perceived easiness to use across the four interfaces. 


\subsubsection{Qualitative responses} 
\label{subsection:6.3.2}
\textbf{Preference}. In the open-response questions after four learning sessions, fifteen participants indicate their preferences for the Story-AI interface for vocabulary learning. 
They especially favor adaptive writing assistance ($N = 9$), generative stories ($5$), and useful practices ($4$). 
\textit{``It (Storyfier-AI) not only provides the meaning, pronunciation, and example sentences of words, but more importantly, it has AI-generated short stories that can help me better understand the meanings of words and how to use them. In addition, the following cloze test and story writing practice can further consolidate my understanding. When I do not know how to write, AI will also provide prompts to help me find my weak points and mistakes so that I can pay more attention on them later on''} (P22).
Six participants prefer the Story-sen interface, and three of them credit the writing practice without AI assistance. \textit{``I prefer Storyfier-sen as I need to rely on myself to think and write down the story, which would be more impressive for vocabulary learning''} (P8).
Five participants prefer the Read-AI interface for its low task workload ($N = 3$), meaningful contexts for learning ($N = 3$), and enjoying the experience ($2$), while the rest two participants favor the Read-sen one as they are more used to the rote learning practices. 

\textbf{Generative stories}. 
Regarding participants' comments on the generative stories, we found positive opinions that they are coherent ($8$) and novel/interesting/impressive ($9$). 
P26 gives us an example. 
\textit{``At first, I could not remember the word `veil'. Then, I checked the generated story, which tells that a veil blocked my vision when I was driving in traffic. This story is close to real life, and I felt terrified. It is impressive. I remembered the word `veil' now.''} 
Nevertheless, there are six comments suggesting that the stories were not coherent, which may be due to the lack of semantic connections among the target words. 
\textit{``When the five target words, \eg `hasten, infinity, jet, basin, and trolley' are not naturally relevant to each other, it would be hard to have a reasonable story that covers them, making it hard for memorizing the words in a batch''} (P27). 
Besides, three users comment that some target words in the generated stories have different meanings from the dictionary ones, and another three users mention that the stories contain some words that are unknown, which disturbs story comprehension. 

\textbf{Cloze test}. 
Twenty users indicate their preferences on using generative stories for the cloze test, which can \textit{``make it easy to connect the words in context''} ($N = 10$),  \textit{``enhance memory of target words''} ($7$), and \textit{``train reading comprehension skills''} ($2$). 
The other eight participants, however, prefer the existing sentences provided by Storyfier-sen for cloze test materials, with four comments mentioning that \textit{``separate sentences are easier to understand''}. 

\textbf{Writing practice}. 
There are seventeen positive responses on the adaptive writing assistance from the generative AI models, suggesting that it can encourage writing ($8$), reduce writing workload ($5$), and provide example usage of target words for reference ($4$). 
\textit{``I didn't feel confused when writing with Storyfier-AI. I can write a sentence first and then let the AI write the next one, and so on. It's like having a buddy to memorize words together, which is more interesting and not boring''} (P2). 
\textit{``The AI's prompts inspire my writing exercises''} (P23). 
However, these prompts in the turn-taking writing process may not match the learners' idea flow ($2$) and language styles ($1$) and cause their reliance on AI for using target words ($1$). 

In all, these qualitative responses reveal 
\pzh{that participants} generally favor Storyfier with AI generative models in Read-AI and Storyfier-AI learning sessions compared to the Read-sen and Storyfier-sen interfaces. However, these models still need to be improved and customized regarding the coherence, complexity, and style of the generative content.
\section{Discussion}

\subsection{Insights from our findings}
\subsubsection{Text generation models for vocabulary learning support} 
In Phase 1, we develop a story generation model and verify that it can generate comparably good stories with the human-written ones given target word sets and titles. 
We receive divided opinions on the generative stories regarding their coherence and interestingness in the experiment with 28 ESL learners. 
The main reason could be that if the target five words are not naturally relevant to each other, it would be difficult for the generative model to connect them to form a meaningful story. 
Besides, we get feedback from English teachers in Phase 1 that our generative stories have generally acceptable complexity for vocabulary learners. 
Yet, there are still cases that our stories contain unknown words in addition to the target ones. 
%

\subsubsection{Vocabulary learning activities}
In Phase 1, we propose that our generative models can empower the cloze test and writing practices in addition to the traditional reading activities that previous vocabulary learning tools support. 
Our evaluation study with ESL learners reveals that they have significantly more learning gains in read-cloze-write \pzh{(\ie Storyfier-sen and Storyfier-AI)} learning sessions compared to that in read-only \pzh{(\ie Read-sen and Read-AI)} sessions.
This is non-surprising as learners spent more effort in understanding target words and practicing their usage in the read-cloze-write sessions. 
Participants' responses in the open-ended questions suggest that generative models can facilitate vocabulary learners by providing meaningful reading and cloze-test materials and adaptive prompts in writing practices. 

\subsubsection{Impact of generative models on vocabulary learning}
In the read-only sessions, we observe that the additional AI-generated stories improve learning engagement but do not improve learning gains. 
This does not support the Gu \etal's implication that learning vocabulary in a batch under a coherent context could facilitate recalls of target words  \cite{vocabulary_learning_strategies}. 
One key reason could be the low semantic relevance among some target words in our experiment. 
Gu \etal's implication could still be valid if the selected target words are topically relevant to each other, \eg as organized in a typical language course book. Generative stories can explicitly reveal the words' relationship. 

In the read-cloze-write sessions, we found that the AI support leads to reduced vocabulary learning gains. 
We attribute these results to the amount of effort spent in the writing practices. 
While participants mostly favor the assistance from our generative model, they spent less time in writing and wrote significantly fewer words in the story. 
This provides a lesson that the AI's assistance should encourage necessary effort in vocabulary learning instead of aiming to reduce learners' workload. 

\subsection{Design Considerations}
Based on our findings, we outline three directions for supporting vocabulary learning with generative models. 

\pzh{
\textbf{Support four strands of vocabulary learning activities with generative models}. 
Our results (\autoref{subsection:6.1}) with 28 ESL learners show that \name{} improves learning gains compared to the read-only baselines. 
This improvement can be due to the \name{}'s cloze test and writing activities that integrate the recommended four strands of learning activities \cite{nation2007four} (\autoref{subsection_2.1}). 
We thus recommend that vocabulary learning tools should integrate multiple strands of activities. 
While prior language learning systems, \eg Smart Titles \cite{kovacs2014smart} and EnglishBot \cite{ruan2021englishbot}, have explored vocabulary learning activities like watching videos and speaking to others, they largely leveraged existing learning materials. 
We suggest that generative techniques, such as the story generation model we developed, the text-to-image~\cite{xu2018attngan}, the text-to-video~\cite{li2018video}, and the music generation \cite{music_generation} approaches, can enrich the learning materials and offer in-situ assistance in these vocabulary learning activities. 
}
For example, the learning support tool can generate an image based on the example sentence of each target word to help them understand the word's meaning. It can further offer generated music clips that use this sentence as listening resources. 

\textbf{Provide adaptive learning feedback.} 
\name{} offers feedback on the correctness of cloze test results and grammar of written sentences. 
However, two participants comment that it could offer more adaptive learning feedback. 
For example, it can \textit{``first let the learner draft a story and then assess its quality and usage of target words''} (P27). 
It can further \textit{``recommend how to improve the written story''} (P25).
Previous skills learning tools, such as ArgueTutor \cite{arguetutor}, Persua \cite{xia2022persua} and VoiceCoach \cite{wang2020voicecoach}, and writing support tools like MepsBot \cite{mepsbot_chi20} also adopt a similar feedback flow, which can mitigate disturbance on the practicing process. 
Generative models can offer such learning feedback by generating polished versions of the written story for reference. 
However, based on the teachers' suggestions on our first \name{} prototype (\autoref{fig:system}D-5), the provided feedback should emphasize the vocabulary learning goal; otherwise, learners may chase for other objectives, \eg trigram repetition and sentence coherence.  

\textbf{Balance machine and human effort in learning tasks}. 
\pzh{Our results show that \name{}'s AI features reduce learning gains on the retention of target words' meanings in the read-cloze-write sessions (\autoref{subsection:6.1}). 
Participants report the need for increased workload and autonomy in the writing task for effective vocabulary learning (\autoref{subsection:6.3.2}). 
As such, we recommend that \name{} should further motivate necessary user effort in learning. 
In the refinement of \name{} (\autoref{subsec.refined_sys}), we have experienced a few features to encourage more user effort, \eg users need to write the first sentence of the story. 
Future work could explore the inclusion of gamification features like badges, timers, and leader boards for promoting learners' efforts in educational scenarios \cite{denny2018empirical}.} 

\subsection{Generality of \name{}}
While \name{} presets the target words that participants are unknown in the experiment, we have equipped it with features like adding or deleting any words, suggesting words semantically related to the title, filtering words based on difficulty level, and editing the generated stories. 
In other words, \name{} can support customized individual vocabulary learning beyond the controlled lab sessions. 
Besides, our English teachers in the design workshop express their interest in applying \name{} in language teaching. 
One teacher, E2, had a trial on her offline course by inputting five words she just taught and asking students to have a cloze test on the generated story. 
Therefore, our \name{} is promising for supporting customized vocabulary learning and teaching in the wild.

\subsection{Limitations and Future Work}
\textbf{Handle language ambiguity}.
Currently, our system does not consider language ambiguity when generating stories for target vocabulary. For example, one word can carry multiple meanings (\ie polysemy) in different contexts, while our system only considers its most common meaning in practical usage. Comparative studies of the same word in different contexts can help disambiguate words and deepen the understanding of vocabulary.

\xingbo{\textbf{Improve story generation quality}.
To further improve the quality of story generation for vocabulary learning, we can consider two aspects: dataset and model.
The simplicity of the ROCStory dataset, while appreciated by English teachers for vocabulary learning, has certain limitations. It lacks transition words, which makes it challenging for models to learn sentence transition logic. Additionally, the simplicity of the story structures can potentially compromise the richness of intra-sentence contexts.
In the future, we can incorporate more complex stories with a wide range of narrative structures into our dataset for model training. 

Besides, we can investigate the use of more powerful language models (\eg ChatGPT) to enhance the quality of the generated stories. 
For example, we have experimented with prompts (\eg with ``simple'', ``CET-4'', ``within 50 words'', and/or ``no more complex words'') to steer ChatGPT towards generating stories that fulfill our specifications  \footnote{For example, we queried ChatGPT using ``write a five-sentences simple story using words: hasten, infinity, jet, basin, and trolley''. This results in a 71-word coherent story but contains more complex sentence structure and words like ``marvel'', ``exhilarated'', and ``adventure''.}.
Notably, however, there is a trade-off between the simplicity and coherence of the generated stories. Future research can focus on refining prompting strategies to optimize the balance between these two elements for enhancing the efficacy of vocabulary learning.}

\textbf{Generate diverse and harmless stories}.
\name{} presents one story at a time based on the generation models trained on ROCStory dataset, which contains simple and short stories. 
In the future, we can consider generating more diverse stories (\eg in the form of newspapers, novels, poems, and humor) that have varied styles and lengths for vocabulary learning.
Besides, while our participants did not report harmful content in the generated stories, \name{} should include features like ``report'' and automatic detection of unwanted content to offer a healthy learning environment.  

\pzh{
\textbf{Identify helpful characteristics of stories for vocabulary learning}.
We evaluate the quality of our generated stories via coherence, relevance, and interestingness (\autoref{table:human_rating}) and collect qualitative feedback from participants on the stories' helpfulness. 
We call for future work to complement our evaluation studies by identifying what are the helpful characteristics of stories for vocabulary learning.
For example, we can talk to English-learning textbook authors or conduct a content analysis on textbook stories. 
The identified characteristics can further guide us to customize story generation models and enhance our evaluation metrics of the stories.}

\textbf{Invite diverse language learners}.
We conduct the experiment with Chinese students in an English learning course to evaluate \name{}. 
Their CET-4 test scores and self-reported proficiency indicate that they are intermediate-level English learners. 
Further studies can explore whether and how \name{} with generative models can support novices with no or little prior experience in learning English.
Moreover, we can extend \name{} to support users from different cultures to learn their foreign languages (\eg English learners study Chinese).

\textbf{Evaluate cloze test and writing practices separately}. 
In the experiment, we evaluate the impact of the cloze test and writing practices by comparing the participants' performance and experience in read-only and read-cloze-write sessions. 
This study design is to identify the value of the vocabulary learning activities beyond the meaning-focused input activities that previous systems support. 
However, we can not quantitatively tell how much the cloze test or writing practice contributes to the impact, which requires a future study that separately evaluates these two activities. 





\section{Conclusion}


In this paper, we designed and developed an interactive system, \name{}, to support reading, cloze test, and writing activities for vocabulary learning.
We power the system with controllable language models that can generate stories given any target words and provide adaptive assistance when using these words in the writing practices. 
We explore its supported vocabulary learning activities and interface design with teachers, learners, and Human-Computer Interaction researchers. 
Our two-by-two within-subjects experiment with 28 English-as-Second-Language Chinese students shows that participants generally favor the generated stories and writing assistance. 
However, their learning gains with \name{} in the read-cloze-test sessions decrease compared to the cases they are with a baseline system without generative models. 
We discuss insights from our findings for leveraging generative models to support learning tasks.

\begin{acks}
This work is supported by the Young Scientists Fund of the National Natural Science Foundation of China with Grant No. 62202509 and partially supported by the Research Grants Council of the Hong Kong Special Administrative Region under General Research Fund (GRF) with Grant No. 16203421. 
\end{acks}

\bibliographystyle{ACM-Reference-Format}
\bibliography{main}

\end{document}